\documentclass[manuscript]{aastex}

\usepackage{natbib}
\usepackage{textcomp}
\usepackage[colorlinks=true,citecolor=blue]{hyperref}
\usepackage{graphicx,color}
\usepackage{dcolumn}
\usepackage{times}
\usepackage{amssymb}

\newcommand{\fermi}{{\it Fermi}}
\newcommand{\phflux}{\mbox{${\rm \, ph \,\, cm^{-2} \, s^{-1}}$}}

\slugcomment{submitted to ApJ}

\shorttitle{The High Redshift Blazar J0809+5341}
\shortauthors{Paliya et al.}

\begin{document}

\title{Awakening of The High Redshift Blazar CGRaBS J0809+5341}

\author{Vaidehi S. Paliya$^{1,\,2}$, M. L. Parker$^{3}$, C. S. Stalin$^{1}$, A. C. Fabian$^{3}$, S. Ramya$^{4}$, S. Covino$^{5}$, G. Tagliaferri$^{5}$, S. Sahayanathan$^{6}$, and C. D. Ravikumar$^{2}$} 
\affil{$^1$Indian Institute of Astrophysics, Block II, Koramangala, Bangalore-560034, India}
\affil{$^2$Department of Physics, University of Calicut, Malappuram-673635, India}
\affil{$^3$Institute of Astronomy, Madingley Road, Cambridge CB3 0HA, UK}
\affil{$^4$Shanghai Astronomical Observatory, 80 Nandan road, Shanghai-200030, China}
\affil{$^5$INAF$-$Osservatorio Astronomico di Brera, Via Bianchi 46, I$-$23807 Merate, Italy}
\affil{$^6$Astrophysical Sciences Division, Bhabha Atomic Research Centre, Mumbai-400085, India}
\email{vaidehi@iiap.res.in}

\begin{abstract}

CGRaBS J0809+5341, a high redshift blazar at $z = 2.144$, underwent a giant optical outburst on 2014 April 19 when it brightened by $\sim$5 mag and reached an unfiltered apparent magnitude of 15.7 mag. This implies an absolute magnitude of $-$30.5 mag, making it one of the brightest quasars in the Universe. This optical flaring triggered us to carry out observations during the decaying part of the flare covering a wide energy range using the {\it Nuclear Spectroscopic Telescope Array}, {\it Swift}, and ground based optical facilities. For the first time, the source is detected in $\gamma$-rays by the Large Area Telescope onboard the {\it Fermi Gamma-Ray Space Telescope}. A high optical polarization of $\sim$10\% is also observed. Using the Sloan Digital Sky Survey spectrum, accretion disk luminosity and black hole mass are estimated as $1.5 \times 10^{45}$ erg s$^{-1}$ and $10^{8.4}~M_{\odot}$ respectively. Using a single zone leptonic emission model, we reproduce the spectral energy distribution of the source during the flaring activity. This analysis suggests that the emission region is probably located outside the broad line region, and the jet becomes radiatively efficient. We also show that the overall properties of CGRaBS J0809+5341 seems not to be in agreement with the general properties observed in high redshift blazars up to now.

\end{abstract}




\section{Introduction}\label{sec:intro}
Blazars are a subclass of active galactic nuclei (AGN) with powerful relativistic jets aligned close to the line of sight to the observer \citep{1995PASP..107..803U}. They emit over the entire electromagnetic spectrum, predominantly by non-thermal emission processes. Because of small inclination angle, the emission from their jet is Doppler boosted. Blazars are classified as flat spectrum radio quasars (FSRQs) and BL Lac objects 
based on the rest frame equivalent width (EW) of their broad optical emission lines with FSRQs having EW $> 5$ \AA~\citep{1991ApJS...76..813S,1991ApJ...374..431S}. However, \citet{2011MNRAS.414.2674G} \citep[see also][]{2009MNRAS.396L.105G} have recently proposed a new classification scheme based on the luminosity of broad line region (BLR) measured in units of Eddington luminosity with FSRQs having $L_{\rm BLR}/L_{\rm Edd} > 5 \times 10^{-4}$. Both classes share many common properties, such as flat radio spectra ($\alpha_r~< 0.5;~\rm{S_{\nu}}~\propto~\rm{\nu}^{-\alpha}$) at GHz frequencies, rapid flux and polarization variations \citep{1995ARA&A..33..163W,2005A&A...442...97A} and exhibit superluminal patterns at radio wavelengths \citep{2005AJ....130.1418J}.

The broadband spectral energy distribution (SED) of blazars consists of two distinct peaks. The low energy peak lies between the regimes of infrared (IR) and X-ray while the high energy one lies in the MeV$-$TeV range. The low energy component is known to result from synchrotron emission whereas the origin of the high energy component is still a matter of debate. In the leptonic emission model, the high energy peak in the SED is explained by the inverse-Compton (IC) scattering of synchrotron photons from the jet \citep[synchrotron self Compton or SSC,][]{1981ApJ...243..700K,1985ApJ...298..114M,1989ApJ...340..181G}. Alternatively, the seed photons for IC scattering can be external to the jet \citep[external Compton or EC,][]{1987ApJ...322..650B,1989ApJ...340..162M,1992A&A...256L..27D}. The high energy component can also be explained as a result of hadronic processes \citep[see e.g.,][]{2003APh....18..593M,2013ApJ...768...54B}. FSRQs and BL Lac objects are found to follow the so-called blazar sequence \citep{1998MNRAS.299..433F,1998MNRAS.301..451G}. However, recently \citet{2012MNRAS.420.2899G} have proposed that the existence of such sequence could be due to selection effects.

CGRaBS J0809+5341 (hereafter J0809+5341) is a high redshift FSRQ \citep[$z = 2.144 \pm 0.002$,][]{2014A&A...563A..54P} which was overlooked for a long time due to its faintness and/or prolonged quiescence. It was predicted as a candidate $\gamma$-ray emitter by \citet{2008ApJS..175...97H}, but was not detected by any earlier $\gamma$-ray surveys. It is radio bright \citep[$F_{\rm 8.4~GHz} = 183.4$ mJy,][]{2007ApJS..171...61H}, having a flat radio spectrum and exhibits compact core morphology in the Faint Images of the Radio Sky at Twenty centimeters (FIRST) observations. 

In this work, we present a detailed multi-wavelength study of J0809+5341 which recently flared in the optical band. By analyzing both archival and simultaneous observations, we present a consistent picture of its emission processes in the framework of leptonic radiation models of blazars in its low and high activity states. In Section~\ref{sec:motiv}, we describe the multi-wavelength campaign carried out to study the optical outburst of this source. The details of the data reduction procedure are reported in Section~\ref{sec:data_red} and the results are presented in Section~\ref{sec:results}. We discuss our findings in Section~\ref{sec:dscsn} and provide conclusions in Section~\ref{sec:conclu}. Throughout the work, we adopt a $\Lambda$CDM cosmology with the Hubble constant $H_0=71$~km~s$^{-1}$~Mpc$^{-1}$, $\Omega_m = 0.27$, and $\Omega_\Lambda = 0.73$.

\section{Optical Outburst And Multi-Wavelength Campaign}\label{sec:motiv}
\begin{figure*}

\hbox{\hspace{-1.5cm}
      \includegraphics[width=10.0cm]{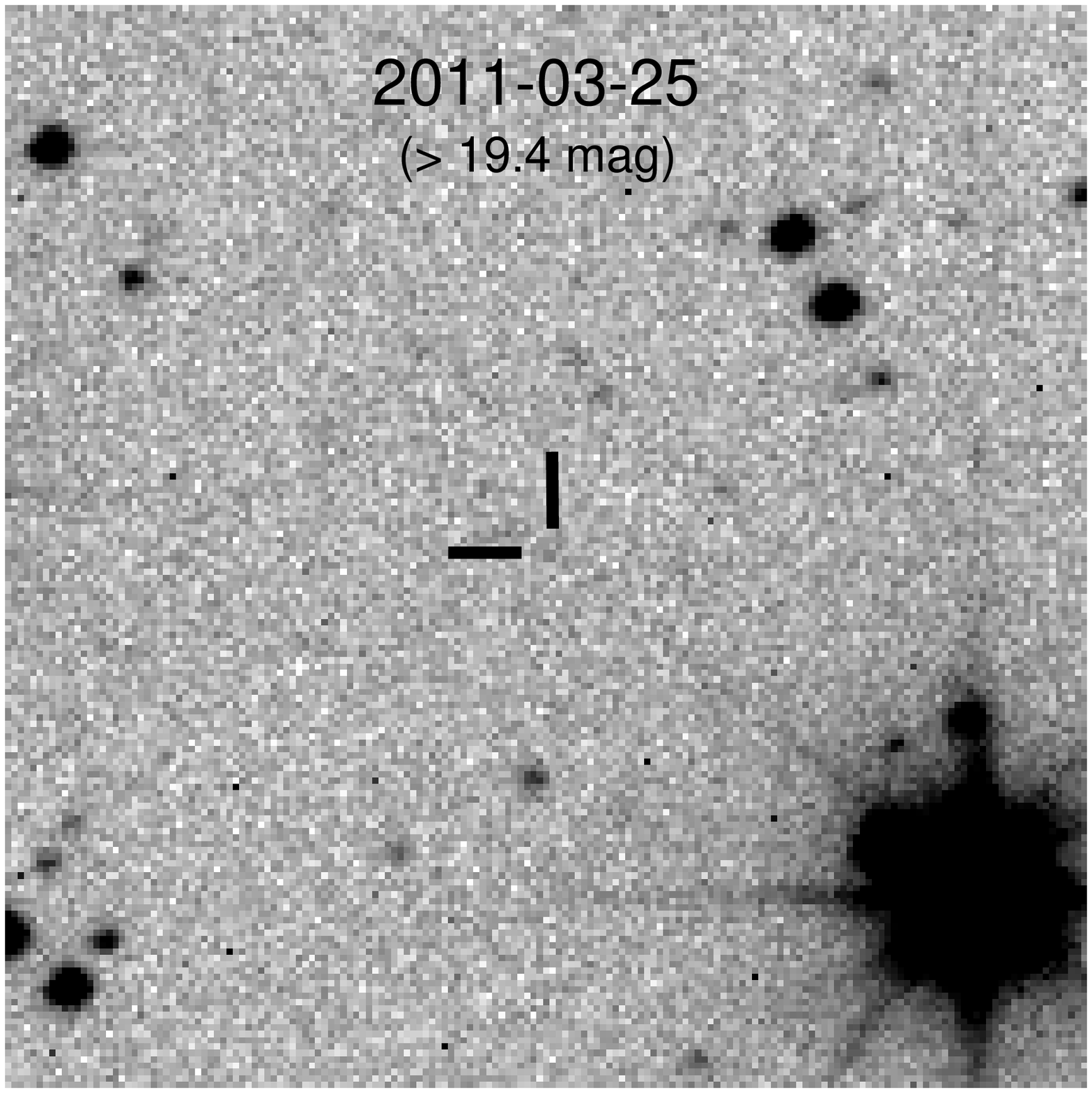}
      \includegraphics[width=10.0cm]{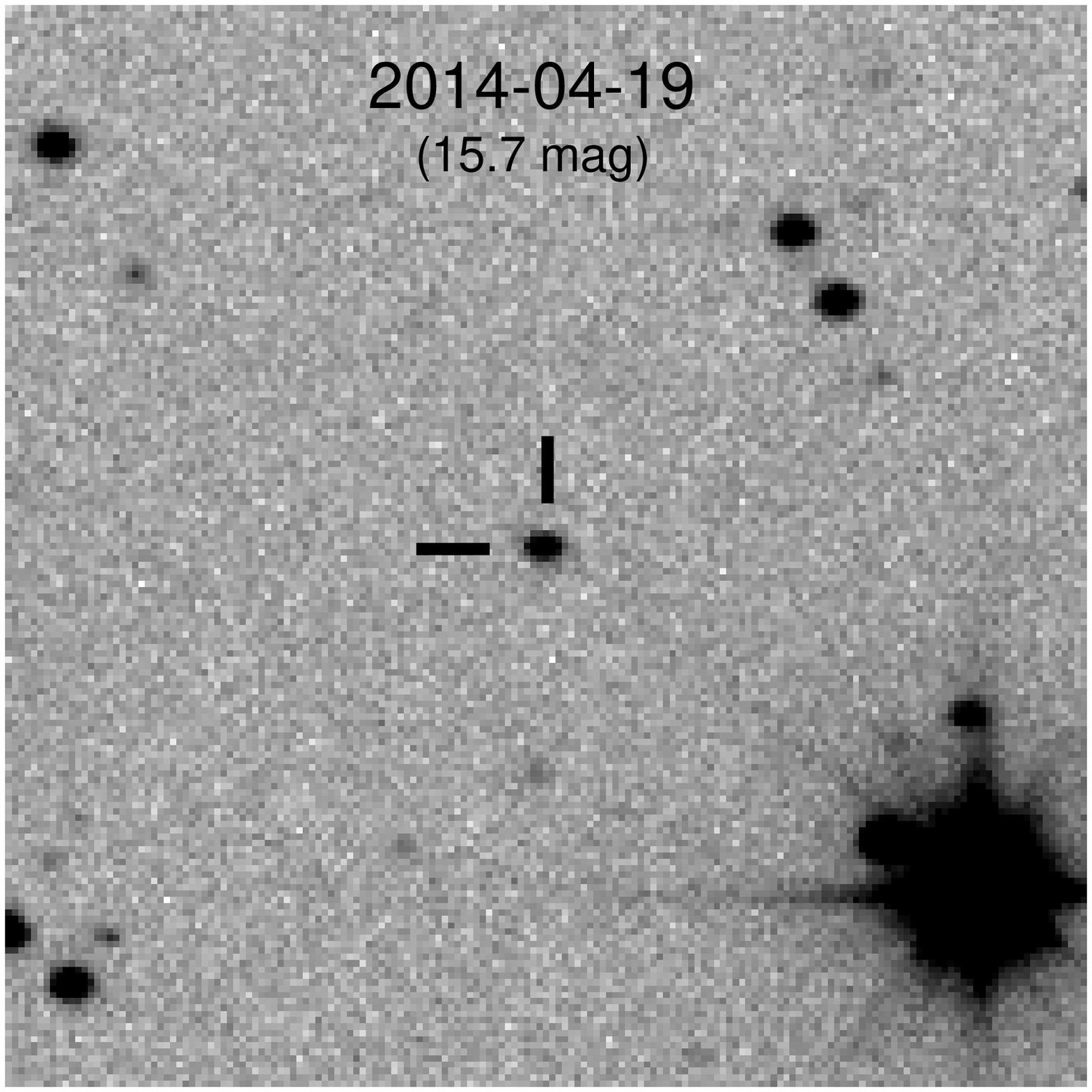}
     }
\caption{Mobile Astronomical System of the TElescope-Robots (MASTER) image of J0809+5341 (field of view of $5^{\prime\prime}\times5^{\prime\prime}$) taken on 2011 March 25 (left) and on 2014 April 19 (right). An upper limit of 19.4 mag is obtained during 2011 observations, while there is $\sim$ 5 mag brightening (with respect to the SDSS observations) during the 2014 optical outburst. North is up and East is left.}\label{fig:optical_flare}
\end{figure*}

J0809+5341 underwent a giant optical outburst on 2014 April 19 \citep{2014ATel.6096....1B}, when its unfiltered apparent magnitude reached 15.7 mag (see Figure~\ref{fig:optical_flare}). This corresponds to an absolute magnitude of $-$30.5 mag, thereby making it one of the brightest quasars in the Universe. Compared to the archival Sloan Digital Sky Survey (SDSS) observations, the source was $\sim$ 5 mag brighter on 2014 April 19. In order to study this extraordinary event, we organized a multi-frequency campaign using both space and ground based telescopes. The hard X-ray mission {\it Nuclear Spectroscopic Telescope Array} \citep[{\it NuSTAR},][]{2013ApJ...770..103H} observed the target on 2014 May 8 and was supplemented by a simultaneous {\it Swift} target of opportunity (ToO) and optical polarimetric observations from the {\it Telescopio Nazionale Galileo}\footnote{http://www.tng.iac.es} (TNG). Additionally, the source was observed on 2014 April 26, 27, and 28 by other {\it Swift} ToO observations. We observed the source in Bessel B, V, R, and I filters on 2014 April 27 and in Bessel U, B, V, R, and I bands on 2014 May 1 from the {\it Himalayan Chandra Telescope}\footnote{http://www.iiap.res.in/centers/iao} (HCT). The entire campaign was complemented by continuous monitoring from the Large Area Telescope (LAT) onboard the {\it Fermi Gamma-ray Space Telescope} (hereafter \fermi-LAT).

\section{Multiwavelength observations and Data Reduction}\label{sec:data_red}

\subsection{{\it Fermi}-Large Area Telescope Observations}\label{subsec:fermi}
The \fermi-LAT data used in this work were collected over the first 72 months of \fermi~operation, from 2008 August 5 (MJD 54,683) to 2014 August 4 (MJD 56,873). We follow the standard data analysis procedures as mentioned in the \fermi-LAT documentation\footnote{http://fermi.gsfc.nasa.gov/ssc/data/analysis/documentation/}. Events belonging to the SOURCE class and in the energy range 0.1$-$300 GeV are used. A filter ``\texttt{DATA$\_$QUAL>0}'' \&\& ``\texttt{LAT$\_$CONFIG==1}'' is used to select good time intervals and a cut of $100^{\circ}$ is applied on the zenith angle to avoid contamination from the Earth limb $\gamma$-rays. Recently released galactic diffuse emission component gll\_iem\_v05\_rev1.fit and an isotropic component iso\_source\_v05\_rev1.txt are considered as background models\footnote{http://fermi.gsfc.nasa.gov/ssc/data/access/lat/BackgroundModels.html}. The normalization parameter of the background models are kept free during the fitting. The binned likelihood method included in the {\it pylikelihood} library of {\tt Science Tools (v9r33p0)} along with the post-launch instrument response functions P7REP\_SOURCE\_V15 are used in the analysis. 

\begin{figure*}
%\centering
\hbox{\hspace{-1.5cm}
      \includegraphics[width=10.0cm]{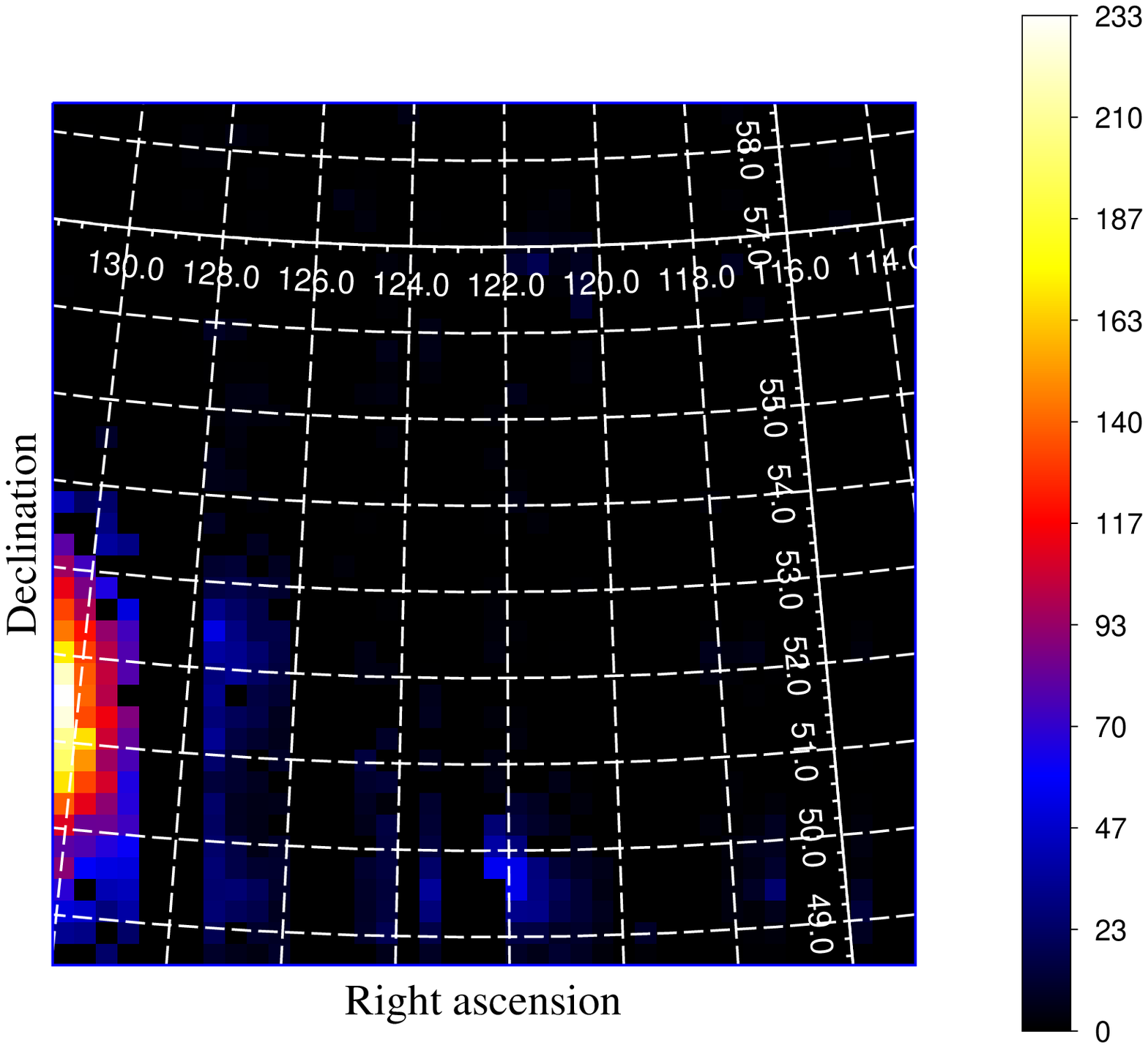}
      \includegraphics[width=10.0cm]{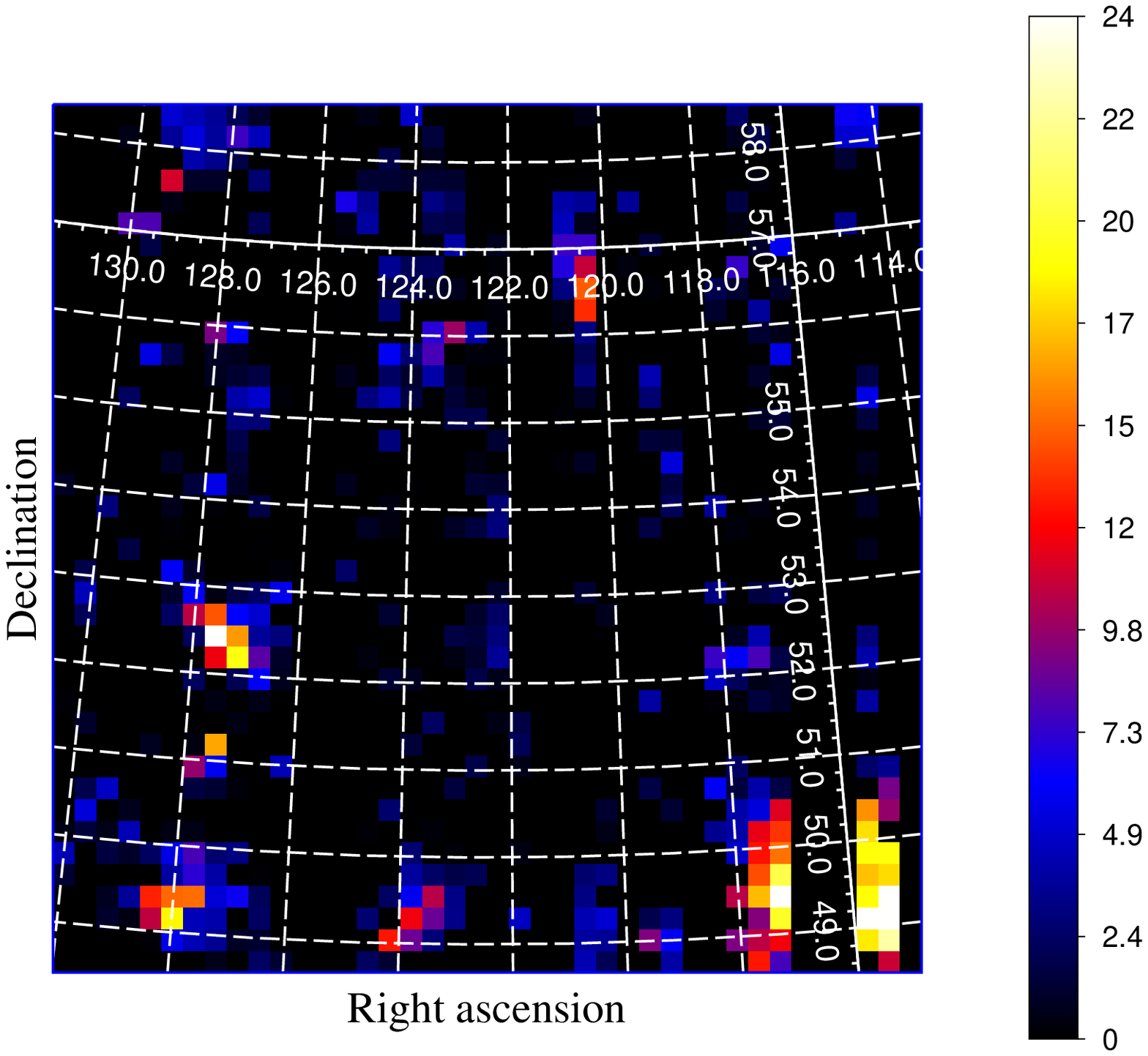}
     }
\caption{Left: Residual TS map of the 0.1$-$300 GeV events during the time period covered in this work, centered on the coordinates of J0809+5341. Two new sources in the ROI are noticed. Right: Residual TS map of the same region after modeling both the new sources.}\label{fig:TSMAP}
\end{figure*}

The significance of the $\gamma$-ray signal is evaluated by means of a maximum likelihood test statistic TS = 2$\Delta \log (\mathcal{L}$) where $\mathcal{L}$ represents the likelihood function, between models with and without a point source at the position of the source of interest. Sources lying within $10^{\circ}$ region of interest (ROI) centered at the position of J0809+5341 and defined in the second \fermi-LAT catalog \citep[2FGL;][]{2012ApJS..199...31N} are included in the model file. The spectral parameters of all the nearby sources are taken from the 2FGL catalog and are allowed to vary (except the scaling factor) during the likelihood fitting. In addition to that, we also include the sources lying between 10$^{\circ}$ to 15$^{\circ}$ from the center of the ROI and keep their parameters fixed to the 2FGL values. Further, \fermi-LAT has detected many sources after the release of the 2FGL catalog and hence these sources are not included in it\footnote{http://www.cbpf.br/icrc2013/papers/icrc2013-1153.pdf}. If lying close to the source of interest, these unmodeled sources could affect the results of the analysis. The presence of these sources, if any, is tested by generating the residual TS map of the ROI. We find two new sources (Figure~\ref{fig:TSMAP}), whose positions are optimized using the tool {\tt gtfindsrc} and obtained as, R.A., Dec. = 122$^{\circ}$.136, 49$^{\circ}$.737 (J2000) and 132$^{\circ}$.506, 51$^{\circ}$.157 (J2000) respectively. The possible counterparts of these $\gamma$-ray sources should be the FSRQ OJ 508 and a $\gamma$-ray emitting narrow line Seyfert 1 galaxy SBS 0846+513 respectively \citep[see][]{2015arXiv150102003T}. We verify that no other significant sources are left in the data to be modeled, by generating another residual TS map including these two new sources in the model file. We do not find any new sources with TS $>$ 25 (see Figure~\ref{fig:TSMAP}). The two new sources are then modeled with a power law model and included in the analysis.

A first round of the likelihood fitting is performed over the entire 72 months of the LAT data and all the sources with TS $<$ 25 are removed from the model. This updated model is then used for further time series and spectral analysis. To generate the spectra and the light curves we allow the normalization parameter of all the sources within the ROI to vary, whereas the photon indices are fixed to the values obtained from the average analysis over appropriate time intervals. Light curves are generated by adopting the unbinned likelihood method as it is expected to encounter fewer events over shorter time intervals\footnote{http://fermi.gsfc.nasa.gov/ssc/data/analysis/scitools/likelihood\_tutorial.html}. The source is considered to be detected if TS $>9$ which corresponds to $\sim 3\sigma$ detection \citep{1996ApJ...461..396M}. For 1 $<$ TS $<$ 9, we calculate 2$\sigma$ upper limit. This is done by varying the flux of the source till TS reaches a value of 4 \citep{2010ApJS..188..405A}. We do not calculate upper limits if TS $< 1$. Primarily governed by uncertainty in the effective area, the measured fluxes have energy dependent systematic uncertainties of around 10\% below 100 MeV, decreasing linearly in log(E) to 5\% in the range between 316 MeV and 10 GeV and increasing linearly in log(E) up to 15\% at 1 TeV\footnote{http://fermi.gsfc.nasa.gov/ssc/data/analysis/LAT\_caveats.html} . Errors associated with the LAT data analysis are the 1$\sigma$ statistical uncertainties, unless otherwise specified.

\subsection{{\it NuSTAR} Observations}\label{subsec:nustar}
The source was observed with \emph{NuSTAR} for a cleaned exposure time of 28~ks. The \emph{NuSTAR} data are reduced and filtered for background flares using the \emph{NuSTAR} Data Analysis Software (NUSTARDAS) version 1.4.1, and response files are generated using CALDB version 20140814. Spectra are extracted for the two focal plane modules (FPMA and FPMB) using the \emph{nuproducts} tool. The source spectrum is extracted from a $30^{\prime\prime}$ circular region centered on the peak emission, and the background spectrum is extracted from a $70^{\prime\prime}$ circular region on the same chip, free of contaminating sources. The spectra are source dominated until $>20$~keV, and they are binned to have a minimum of 20 counts per bin.

\subsection{{\it Swift} Observations}\label{subsec:swift}
The source is below the sensitivity limit of the Burst Alert Telescope \citep[BAT;][]{2005SSRv..120..143B}. However, it is significantly detected by the pointed observations from the X-Ray Telescope \citep[XRT;][]{2005SSRv..120..165B} and the Ultraviolet Optical Telescope \citep[UVOT;][]{2005SSRv..120...95R}. This is the first time that J0809+5341 is detected in the X-ray band.

The XRT data are processed with the XRTDAS software package (v.3.0.0) available within the HEASOFT package (6.16). Event files are cleaned and calibrated using standard filtering criteria with {\tt xrtpipeline (v.0.13.1)} and the calibration database that was updated on 2014 July 30. Standard grade selections of 0-12 in the photon counting mode are used. Cleaned event files are then summed using the task XSELECT. To extract the source spectrum from the summed event files, a circular region of 20 pixel ($\sim$47$^{\prime\prime}$) centered at the source position, is chosen, while background is extracted from a nearby circular region of 50 pixel radius. All the exposure maps are combined with XIMAGE and used to generate ancillary response files using the task {\tt xrtmkarf}. The source spectrum is binned to have at least 1 count per bin. Spectral fitting is done using XSPEC \citep{1996ASPC..101...17A}. An absorbed power law \citep[$N_{\rm H} = 3.75 \times 10^{20}$ cm$^{-2}$;][]{2005A&A...440..775K} is used for fitting. Due to the faintness of the source, we use C-statistics \citep{1979ApJ...228..939C} within XSPEC. The uncertainties are calculated at 90\% confidence level. 

UVOT has observed the source in 4 filters namely V, B, U, and UVW1. All the observations are integrated using the task {\tt uvotimsum} and analyzed with {\tt uvotsource}. Source region is selected as a circle of 5$^{\prime\prime}$ radius centered at the source position. Background is chosen from a nearby source free circular region of 1$^{\prime}$ radius. The observed magnitudes are corrected for reddening using the galactic extinction of \citet{2011ApJ...737..103S} and converted to flux units using the zero point magnitudes of \citet{2011AIPC.1358..373B}. The optical-UV flux of the high redshift blazars ($z > 2$) could be significantly absorbed by neutral Hydrogen in the intervening Lyman $\alpha$ absorption systems. We use the attenuation calculated by \citet{2010MNRAS.405..387G} to correct for this effect.

\subsection{Ground Based Optical Observations}\label{subsec:ground}
\subsubsection{Polarimetry}\label{subsubsec:pol}
Linear polarimetry was carried out on 2014 May 8 at the TNG. Observations were performed with the PAOLO polarimeter\footnote{http://www.tng.iac.es/instruments/lrs/paolo.html} with the Sloan r filter for about an hour. Data reduction as well as aperture photometry are performed using custom-made software tools\footnote{https://pypi.python.org/pypi/SRPAstro.FITS}. Instrumental Stokes parameters, polarization degree, and position angle are then corrected for the instrumental polarization by means of a suitable number of polarized and un-polarized standard stars as described in detail in \citet{2014AN....335..117C}. Flux calibration is derived by the observation of the SDSS isolated and non-saturated stars in the polarimeter field of view (see next section). No flux variability, total or polarized, is detected during the observations.

\subsubsection{Photometry}\label{subsubsec:photo}
As mentioned earlier, J0809+5341 was observed using HCT on two epochs, namely 2014 April 27 in  B, V, R, and I filters and on 2014 May 1 in U,B,V, R, and I filters. The details of the instrument can be found in \citet{2014ApJ...789..143P}. Standard procedures in {\tt IRAF} are used to do the pre-processing of the images (bias subtraction, flat-fielding, and cosmic ray removal). After pre-processing, instrumental magnitudes of the target as well as stars in the field are obtained via PSF photometry using DAOPHOT \citep{2011ascl.soft04011S} available in MIDAS\footnote{Munich Image Data Analysis System}. We could not convert these instrumental magnitudes to standard magnitudes directly as Landolt standard star fields were not observed during the observations. Also, there are no stars in the field of the source with UBVRI photometry available. Therefore, to convert the measured magnitudes to the standard system the following procedure is adopted. We obtain ugriz magnitudes for 3 comparison stars in the field from the SDSS\footnote{http://skyserver.sdss3.org/dr10} (SDSS J080938.95+533813.8, SDSS J080956.48+534341.3 and SDSS J081006.16+534248.0). For these three stars the UBVRI magnitudes are derived using the following transformation equations \citep{2006A&A...460..339J} 
\begin{equation}
U = B + (0.52 \pm 0.06) * (u - g) + (0.53 \pm 0.09) * (g - r) - (0.82 \pm 0.04)
\end{equation}
\begin{equation}
B = g + (0.313 \pm 0.003) * (g - r) + (0.219 \pm 0.002)
\end{equation}
\begin{equation}
V = I + (0.675 \pm 0.002) * (g - i) + (0.364 \pm 0.002)
\end{equation}
\begin{equation}
R = I + (0.930 \pm 0.005) * (r - i) + (0.259 \pm 0.002)
\end{equation}
\begin{equation}
I = i - (0.386 \pm 0.004) * (i - z) - (0.397 \pm 0.001)
\end{equation}
Once the UBVRI magnitudes of these three stars are obtained, the derived instrumental magnitudes of J0809+5341 are converted to UBVRI magnitudes using differential photometry. The standard magnitudes are then dereddened and converted to flux using the zero points of \citet{1998A&A...333..231B}. The flux are corrected for absorption by intervening Lyman $\alpha$ absorption systems as done by \citet{2010MNRAS.405..387G}.

Recently, \citet{2014ATel.6136....1C} have reported the NIR observations of J0809+5341 from the 2.1 m telescope of the Guillermo Haro Observatory operated by the National Institute for Astrophysics, Optics and Electronics (Mexico). We take their reported magnitudes in J, H, and K$_{\rm S}$ bands, correct for reddening and convert to flux units using the zero points of \citet{1998A&A...333..231B}.

\section{Results}\label{sec:results}
\subsection{Black Hole Mass And Accretion Disk Luminosity}\label{subsec:bhmass}
\begin{figure*}
\centering
\includegraphics[width=14.0cm]{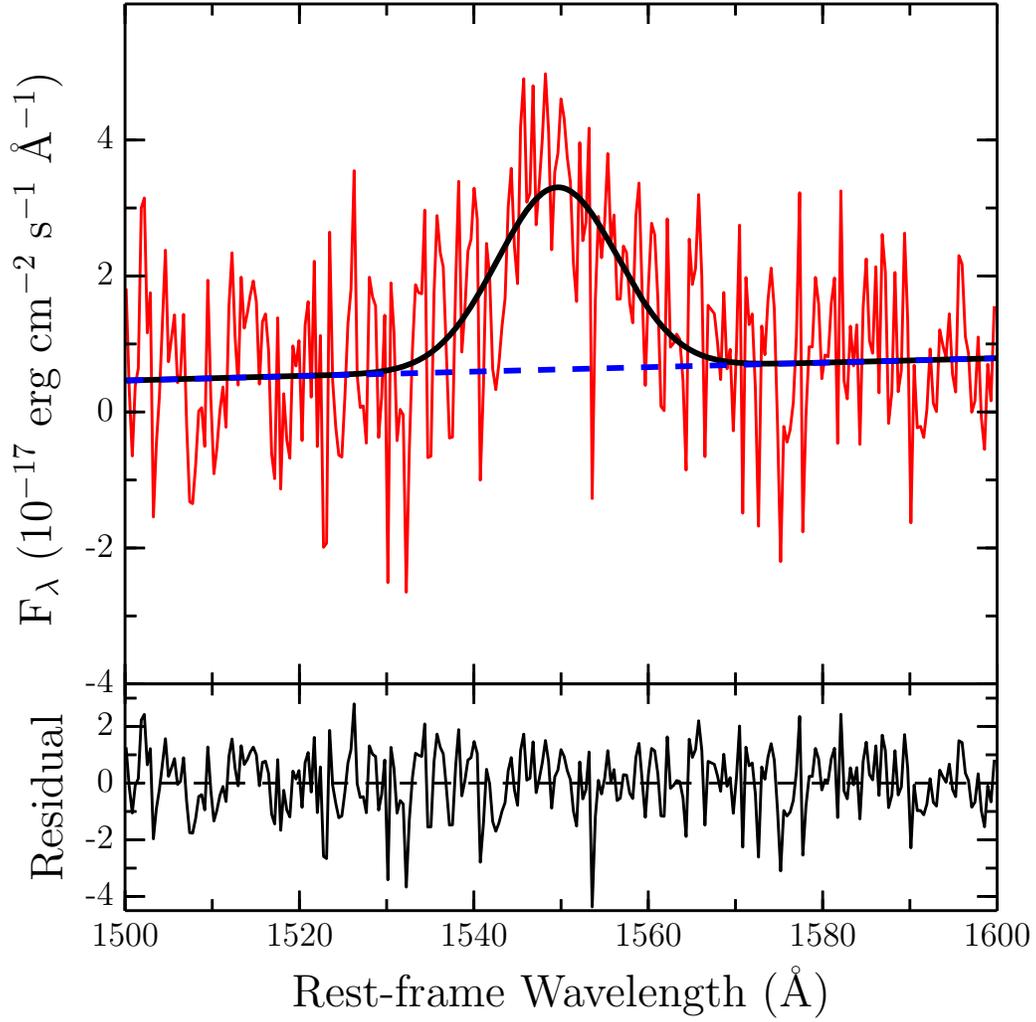}
\caption{Top: A fit to the continuum and C~{\sc iv} line in the SDSS spectrum of J0809+5341. Bottom: The residual between the fit and the observed spectrum.}\label{fig:sdss_fit}
\end{figure*}

J0809+5341 has a low S/N spectrum from SDSS-DR10 \citep{2014A&A...563A..54P}. As this is the only spectrum available for this source, we use it to estimate the black hole mass ($M_{\rm BH}$) and the accretion disk luminosity ($L_{\rm disk}$). The spectrum is brought to the rest-frame and then dereddend using E(B$-$V) of 0.038 taken from the NED\footnote{http://ned.ipac.caltech.edu/}. The fitting to the SDSS spectrum is based on chi-square minimization using the {\tt MPFIT} package \citep{2009ASPC..411..251M}. A  power law continuum is fit to the spectrum using the line free regions [1445, 1465] \AA~and [1700,1710] \AA, i.e. on either side of the C~{\sc iv} line. This continuum is then subtracted from the spectrum. Fe emission is not subtracted from the spectrum as it is known to be weak for C~{\sc iv} line \citep{2011ApJS..194...45S}. The modified spectrum, between the wavelength range [1500, 1600] \AA, is then fit with a single Gaussian function. A narrow component C~{\sc iv} is not considered in our fitting as virial black hole mass estimates using C~{\sc iv} are based on the FWHM of the entire C~{\sc iv} line profile \citep{2006ApJ...641..689V}. The fit to the C~{\sc iv} line is shown in Figure~\ref{fig:sdss_fit}. From the single Gaussian fit we find the line flux and the $\sigma$ of the C~{\sc iv} line as (46.46 $\pm$ 0.76) $\times$ 10$^{-17}$ erg cm$^{-2}$ s$^{-1}$ \AA$^{-1}$ and 1338.3 $\pm$ 139 km s$^{-1}$ respectively. Correcting the observed $\sigma$ for the resolution of the instrument, the FWHM of C~{\sc iv} is estimated as 3145 $\pm$ 327 km s$^{-1}$. Using the region between 1345$-$1350 \AA, we find the mean continuum flux at 1350 \AA~to be (4.96 $\pm$ 0.26) $\times$ 10$^{-17}$ erg cm$^{-2}$ s$^{-1}$ \AA$^{-1}$. This translates to a continuum luminosity ($\lambda L_{\lambda}$) of (2.43 $\pm$ 0.14) $\times$ 10$^{45}$ erg s$^{-1}$. The C~{\sc iv} line luminosity is obtained as $L_{\rm C~{\sc iv}}$ = (1.69 $\pm$ 0.03) $\times$ 10$^{43}$ erg s$^{-1}$. Using the measured FWHM and the continuum luminosity, we estimate the black hole mass using the following equation \citep{2011ApJS..194...45S}
\begin{equation}\label{eqn:virial_estimator}
\log \left({M_{\rm BH,vir} \over M_\odot}\right)
=a+b\log\left({\lambda L_{\lambda} \over 10^{44}\,{\rm
erg\,s^{-1}}}\right)+2\log\left({\rm FWHM\over km\,s^{-1}}\right)\
.
\end{equation}
The adopted values of a and b are 0.66 and 0.53 respectively, taken from \citet{2011ApJS..194...45S}. This results in a black hole mass of log $\left ({M_{\rm BH} \over M_\odot} \right) =  8.39 \pm 0.21$. It is important to note that for jet dominated sources like blazars, single epoch black hole mass estimates may not be relied upon, particularly in the flaring states as (i) the BLR will be ionized by non-thermal continuum emission and (ii) the BLR cannot be considered to be in equilibrium during episodes of strong flaring activity \citep{2013ApJ...763L..36L}. However, since the available SDSS spectrum corresponds to the faint state of the source, the black hole mass estimated here seems robust. Following \citet{1997MNRAS.286..415C} \citep[see also][]{1991ApJ...373..465F}, we calculate the total BLR luminosity ($L_{\rm BLR}$) using the flux of the C~{\sc iv} line. We find $L_{\rm BLR}$ = 1.5 $\times$ 10$^{44}$ erg s$^{-1}$. Assuming 10\% of the accretion disk luminosity is reprocessed by the BLR, the accretion disk luminosity is 1.5 $\times$ 10$^{45}$ erg s$^{-1}$.

\subsection{Average Gamma-Ray Properties}\label{subsec:lat_results}
J0809+5341 is not present in the 2FGL catalog, indicating that it was not detected with TS $> 25$ in first two years of \fermi~operation. Indeed, the LAT data analysis for this period results in TS = 0.5, explaining the absence of the source in the 2FGL catalog. However, analysis of the third through fifth year of the LAT data gives TS $\approx$ 185.5 \citep[$\sim$ 13$\sigma$,][]{1996ApJ...461..396M}, thus confirming that J0809+5341 is a high redshift $\gamma$-ray emitting FSRQ. Therefore, for the first time, we report the detection of J0809+5341 in the $\gamma$-ray band\footnote{This source is now included in the recently released 3FGL catalog as 3FGL J0809.5+5342 \citep{2015arXiv150102003T}.}. Moreover, the source is found to be in a relatively bright state during the sixth year of \fermi~operation with TS = 191.3 and the derived $\gamma$-ray flux and photon index are (2.78 $\pm$ 0.44) $\times$ 10$^{-8}$ \phflux~and 2.15 $\pm$ 0.08 respectively. The details of the results of the average analysis of \fermi-LAT data, covering various time intervals, are given in Table~\ref{tab:fermi_lat}.

We perform $\gamma$-ray point source localization over the photons extracted during the third through fifth year of \fermi~operation (the time period when the source was detected significantly) using the tool {\tt gtfindsrc}. The localized spatial coordinates are RA = 122$^{\circ}$.446, Dec = 53$^{\circ}$.673 (J2000), at an angular separation of 0$^{\circ}$.02 from the radio position of J0809+5341 (RA = 122$^{\circ}$.424, Dec = 53$^{\circ}$.690, J2000), with a 95\% error circle radius of 0$^{\circ}$.04. This implies a close spatial association of the $\gamma$-ray source with the radio counterpart.

In order to search for the energy of the highest energy photon, we use the tool {\tt gtsrcprob} and event class CLEAN. The highest energy is found to be 13.63 GeV, detected on 2013 December 31, at 0$^{\circ}$.03 far from the source position, and with 98.4\% probability to be associated with the $\gamma$-ray source.

\subsection{Gamma-ray Temporal Variability}\label{subsec:mw_var}
\begin{figure*}
\centering
% \hspace{-2.0cm}
 \includegraphics[width=\textwidth]{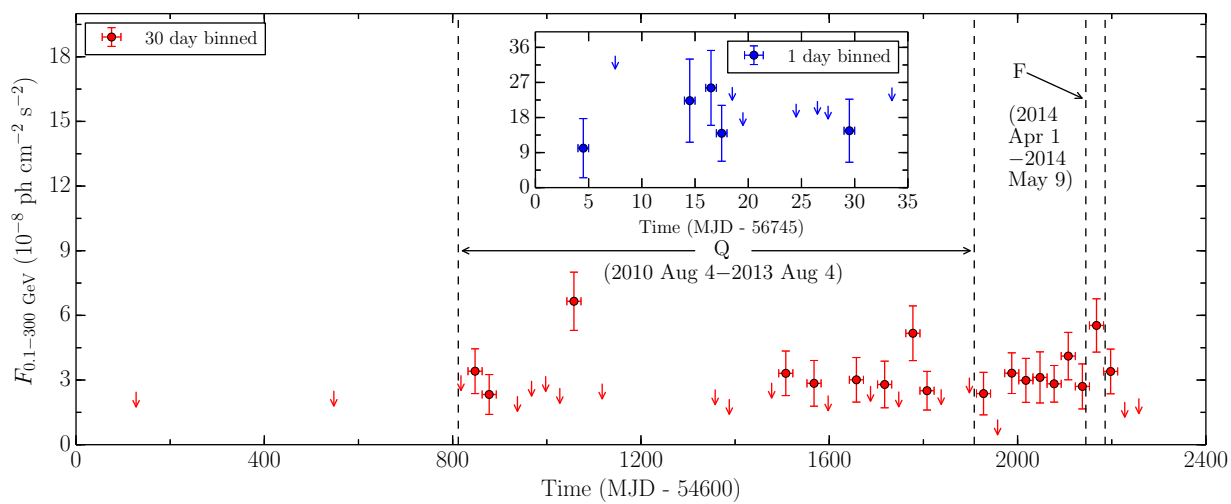}
\caption{Flux history of J0809+5341 in the $\gamma$-ray band, covering the first 72 months of \fermi-LAT operation. Data are in units of 10$^{-8}$ \phflux. 95\% upper limits are shown by downward arrows. Q represents the low activity state, while F denotes the high activity state selected for the SED modeling. Inset: Daily binned $\gamma$-ray light curve covering the period of high activity. The flux units are same as that of the main panel.}\label{fig:long_lc}
\end{figure*}

The $\gamma$-ray light curve of J0809+5341, covering the first 72 months of \fermi~operation, is shown in Figure~\ref{fig:long_lc}. \fermi-LAT data are binned monthly and we also show daily scaled $\gamma$-ray flux covering the period of high optical activity (MJD 56,748$-$56,786 or 2014 April 1 to 2014 May 9), by blue circles in the inset. It is clear from Figure~\ref{fig:long_lc} that the source was not detected at all in the first two years of \fermi~operation. It was detected by the LAT sporadically during the third through fifth year at a low flux level. However, it becomes relatively active only during the last 12 months when it was continuously detected by the LAT. Visual inspection of the daily binned light curve shown in the inset of Figure~\ref{fig:long_lc} hints for the presence of two peaks approximately at the same flux level (one at around MJD 56,760 and other at MJD 56,775). Moreover, we calculate the daily binned $\gamma$-ray flux ($F_{\gamma}$) for the flaring period and find a maximum $F_{\gamma}$ of (2.56 $\pm$ 0.96) $\times$ 10$^{-7}$ \phflux~on 2014 April 14 (MJD 56,761), which coincides with the first report of the high optical activity from the source \citep{2014ATel.6070....1S}.

\subsection{Optical-UV Observations}\label{subsec:hct_tng}
The results of multi epoch {\it Swift}-UVOT observations are given in Table~\ref{tab:uvot_phot}. It is evident from this table that the source has shown flux variations on both a daily and weekly timescales. Also, multi-band apparent magnitudes, as observed from HCT, are presented in Table~\ref{tab:hct_phot}. For a comparison with the archival data, we also give the SDSS magnitudes converted to UBVRI filters using the transformations given by \citet{2006A&A...460..339J}. These data are not corrected for Galactic reddening. It is clear from Table~\ref{tab:hct_phot} that the source has brightened significantly in all the filters compared to the archival observations.

The polarimetric observations from TNG on 2014 May 8 show the detection of high optical polarization from J0809+5341. The observed polarization is found to be as high as 9.8 $\pm$ 0.5\%, which is typically seen in blazars. The corresponding observed polarization angle is 98 $\pm$ 1 deg.

\subsection{Spectral Analysis}\label{subsec:spect}
We test the presence of curvature in the overall $\gamma$-ray spectrum of the source by fitting a LogParabola model. It is defined as { $dN/dE \propto (E/E_{\rm o})^{-\alpha-\beta log({\it E/E_{\rm o}})}$}, where $E_{\rm o}$ is an arbitrary reference energy fixed at 300 MeV, $\alpha$ is the photon index at $E_{\rm o}$ and $\beta$ is the curvature index which defines the curvature around the peak. The test statistic of curvature is then evaluated as ${TS_{\rm curve}} = 2 (\log \mathcal{L}({\rm LogParabola}) - \log \mathcal{L}$(power-law)). We set the threshold ${TS_{\rm curve}} \geq 16$ to test the presence of a significant curvature as done by \citet{2012ApJS..199...31N}. There are hints of the presence of curvature as ${TS_{\rm curve}} \approx 10 $ ($\sim3\sigma$; Table~\ref{tab:fermi_lat}), though a strong claim can not be made as $TS_{\rm curve}$ is below the threshold of 16 set in the 2FGL catalog \citep[][]{2012ApJS..199...31N}.

\begin{figure*}
\centering
%\hspace{-2.0cm}
 \includegraphics{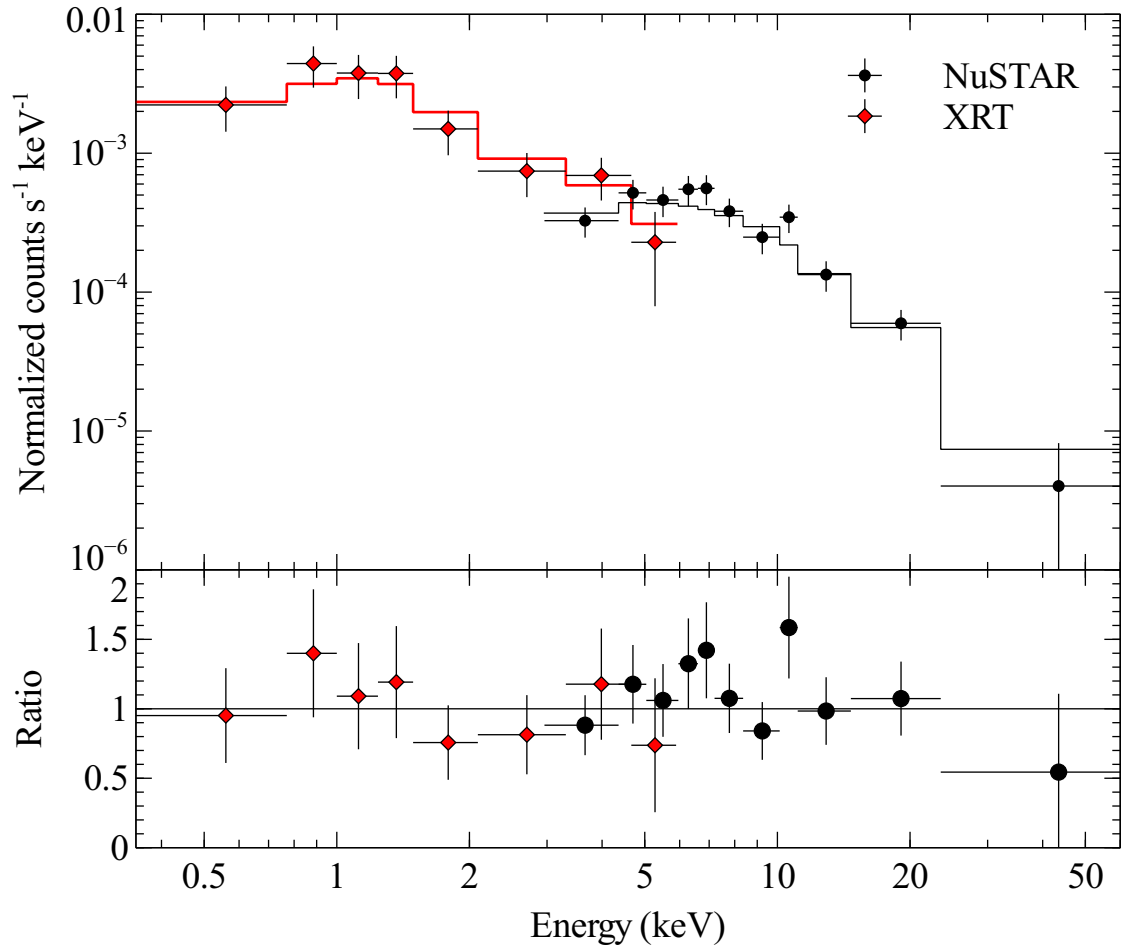}
\caption{Top: Joint XRT (from 0.3--7~keV) and {\it NuSTAR} (from 3--60~keV) spectrum, fit with an absorbed power-law. Bottom: data to model ratio for this fit. Data are rebinned for clarity, and the two \emph{NuSTAR} spectra are grouped in Xspec but fit separately.}\label{fig:xrt_nustar}
\end{figure*}

The joint XRT/\emph{NuSTAR} spectrum is well fit with a simple power law (C-statistic/d.o.f.$=40/35$), modified by Galactic absorption, and we find a best-fit photon index of 1.4 $\pm$ 0.1. This fit is shown in Figure~\ref{fig:xrt_nustar}, along with the residuals to the model. We allow for a difference in flux calibration between {\it NuSTAR} and XRT spectra by including a constant (CONST in XSPEC) fixed at 1 for two {\it NuSTAR} spectra \citep[calibration difference between FPMA and FPMB are on the order of 1\%, and are not detectable in spectra with few counts, see e.g.,][]{2014ApJ...787...83M} and free to vary for the XRT. This constant offset is consistent with 1 (1.2 $\pm$ 0.3) and we find the same value of photon index whether or not we allow for this constant multiplicative factor. The large error is caused by the low count rate, and low sensitivity of the XRT compared to other X-ray telescopes. We calculate the goodness of the fit using Monte-Carlo simulations with the GOODNESS command in XSPEC, and find that ~47\% of the simulated spectra based on the model have a lower $\chi^{2}$.

Fitting the absorbed power law model to the combined XRT spectra of the first three observations on 2014 April 26, 27 and, 28 (for a total exposure of 4.8 ksec yielding 48 counts) resulted in a photon index of 1.2 $\pm$ 0.4 while that derived for the fourth observation on 2014 May 8 (for a total exposure of 4.6 ksec yielding 40 counts) is 1.9 $\pm$ 0.5. This suggests a possible softening of the X-ray spectrum over a course of $\sim$10 days. However, considering the large errors in the photon indices, due to the faintness of the source, a firm conclusion cannot be reached.

The B$-$R color of the source, from HCT observations, are found to be 0.90 $\pm$ 0.04 and 0.94 $\pm$ 0.05 for the epochs of 2014 April 27 and 2014 May 1 respectively and thus there is no color change between these two epochs. At the epoch of the SDSS observation, we find B$-$R = 0.98 $\pm$ 0.06. Thus, we do not see any optical color variation between the epochs 2003 and 2014, though the source has varied significantly in flux.

\subsection{Spectral Energy Distributions}\label{subsec:sed}
We generate the SED of the source during two different activity states. A high activity state covering the period of the recent optical outburst (2014 April 1 to 2014 May 9), and a low activity state for which the LAT data covering the third through fifth year of \fermi~operation (see Figure~\ref{fig:long_lc}) and other non-simultaneous archival observations\footnote{http://www.asdc.asi.it/bzcat} are used. The derived flux values are given in Table~\ref{tab:sed_par}. Assumption of the archival observations as low activity state can be justified by the fact that during the recent optical flare, the optical magnitude was brighter by $\sim$ 5 mag compared to the archival SDSS measurements \citep{2014ATel.6096....1B}. However, since the SDSS observations were taken well before the launch of \fermi~satellite, the results based on such non-simultaneous data can be questioned. The source was observed by Mobile Astronomical System of the TElescope-Robots \citep[MASTER;][]{2010AdAst2010E..30L} on 2011 March 25, i.e. between the third to fifth year of \fermi~operation (see Figure~\ref{fig:optical_flare}). The obtained upper limit in the unfiltered magnitude was 19.4 mag. This hints that during the third through fifth year of \fermi~operation, the optical flux level of the source was possibly similar to that observed by the SDSS in 2003. Moreover, as can be seen in Figure~\ref{fig:long_lc}, the source become active only very recently, therefore the third through fifth year of \fermi-LAT observations can be adopted as low activity state.

We use a simple one zone leptonic emission model, similar to the one used by \citet[][hereafter GT09]{2009MNRAS.397..985G} and \citet{2008ApJ...686..181F} \citep[see also][]{2009ApJ...692...32D}, to interpret the SEDs of the source. The emission region is assumed to be spherical, filled with relativistic electrons, located at a distance $Z_{\rm diss}$, from the central black hole of mass $M_{\rm BH}$, and moving relativistically with bulk Lorentz factor $\Gamma$. A scaling factor of $\Gamma \propto Z^{0.5}$ is assumed (where $Z$ is the distance from the black hole) in the inner parts of the jet where it is anticipated to be accelerating and parabolic in shape \citep[GT09,][]{2004ApJ...605..656V}. After the acceleration phase, the jet becomes conical with semi aperture angle 0.1 rad and a constant bulk Lorentz factor $\Gamma_{\rm max}$.

The electron energy distribution is assumed to follow a smoothly joining broken power law of the form
$N'(\gamma')  \, = \, N'_0\, { (\gamma'_{\rm b})^{-p} \over
(\gamma'/\gamma'_{\rm b})^{p} + (\gamma'/\gamma'_{\rm b})^{q}}$
,where $p$ and $q$ are the particle indices before and after the break energy ($\gamma'_{\rm b}$) respectively (primed quantities are measured in the comoving frame). The size of the emission region is adopted by considering it to cover the entire jet cross-section. Thermal emission from the accretion disk is evaluated assuming a standard optically thick, geometrically thin disk \citep{1973A&A....24..337S} with inner and outer radii $R_{\rm in} = 3R_{\rm Sch}, R_{\rm out} = 500R_{\rm Sch}$ respectively (GT09), where $R_{\rm Sch}$ is the Schwarzschild radius. Locally, the accretion disk spectrum is explained by a multi-temperature blackbody \citep[e.g.][]{2002apa..book.....F}. Above and below the accretion disk, the presence of the X-ray corona is also considered which reprocesses 30\% of the accretion disk luminosity. The inner and outer radii of the corona are assumed to be $3R_{\rm Sch}$ and $30R_{\rm Sch}$ respectively. The spectrum emitted by the corona is adopted as cut-off power law: $L_{\rm cor}(\epsilon)\propto \epsilon^{-\alpha_{\rm cor}}\exp(-\epsilon/\epsilon_{\rm c})$ (GT09), where $\epsilon$ is the dimensionless photon energy (= $\frac{h\nu}{m_{e}c^2}$). The cut-off energy is assumed to be 150 keV and we adopt a flat power law i.e. $\alpha_{\rm cor} = 1$. The BLR is assumed as a spherical shell located at a distance $Z_{\rm BLR} = 10^{17} L^{1/2}_{\rm d,45}$ cm, where $L_{\rm d,45}$ is the accretion disk luminosity in units of $10^{45}$ erg s$^{-1}$ (GT09). It reprocesses 10\% of the accretion disk luminosity and its spectrum is assumed to be a blackbody peaking at rest-frame Lyman-$\alpha$ frequency \citep{2008MNRAS.386..945T}. The dusty torus, for simplicity, is assumed to be a thin spherical shell, located at a distance of $Z_{\rm IR} = 10^{18} L^{1/2}_{\rm d,45}$ cm and reprocess 50\% of the accretion disk emission. Its emission profile can be explained by a simple blackbody peaking at temperature $T_{\rm IR}$. In the presence of randomly oriented magnetic field, electrons radiate via synchrotron and IC scattering mechanisms. The synchrotron and synchrotron self-Compton emissions, in the observer's frame, are calculated using the formulations of \citet{2008ApJ...686..181F}. Besides these mechanisms, the particle also loose energy through IC scattering of external photon field from the accretion disk (EC-disk), BLR (EC-BLR), and the dusty torus (EC-torus) \citep[GT09;][]{2009ApJ...692...32D,2009herb.book.....D,2013MNRAS.436.2170C}. Finally, the kinetic power of the jet is calculated assuming both protons and electrons to have equal number densities \citep{1997MNRAS.286..415C}. Protons are assumed to be cold and contribute only to the inertia of the jet. For the SED modeling, we adopt the black hole mass as 10$^{8.4} M_{\odot}$ and the accretion disk luminosity as 1.5 $\times$ 10$^{45}$ erg s$^{-1}$, as calculated in Section~\ref{subsec:bhmass}. Since the SED in the low activity state is generated using non-simultaneous data, we perform the modeling only on the flaring state SED where the source was monitored contemporaneously over a wide energy range. In Figure~\ref{fig:sed}, we show the model spectrum due to different emission mechanisms along with the observed fluxes and the corresponding parameters are given in Table~\ref{tab:sed}. The variation of the radiation energy densities, measured in the comoving frame, are also shown in the bottom panel of Figure~\ref{fig:sed}. To model the SEDs, we start with a plausible set of parameters which are then constrained within a range that better represent the data.

\section{Discussion}\label{sec:dscsn}
The non-detection of J0809+5341 by earlier high energy missions indicates that either the source is intrinsically faint or remains in quiescence for a long time. The recent optical outburst along with the increased brightness across the electromagnetic spectrum made it possible to detect the source for the first time in X-rays. This also led to the first detection of the source in the $\gamma$-ray band as predicted earlier by \citet{2008ApJS..175...97H}.

In Figure~\ref{fig:optical_flare}, we show the MASTER image of J0809+5341 taken on 2011 March 25 and 2014 April 19. As can be seen clearly, the source was not detected in 2011 and only an upper limit in the unfiltered magnitude was obtained. However, it becomes extremely bright in 2014. The $\gamma$-ray light curve also shows that J0809+5341 became active only in the sixth year of \fermi~operation when enhanced $\gamma$-ray emission is observed during the optical outburst. Comparing the results of the $\gamma$-ray analysis done from the third through fifth year of \fermi~operations with that obtained during the sixth year reveals that (i) the flux increased in the sixth year, and (ii) the photon index does not change within the errors.

During the period of high activity, the maximum one day binned $\gamma$-ray photon flux is found to be (2.56 $\pm$ 0.96) $\times$ 10$^{-7}$ \phflux~which corresponds to an isotropic $\gamma$-ray luminosity ($L_{\gamma}$) of 9.3 $\times$ 10$^{48}$ erg s$^{-1}$. This, in turn, corresponds to a luminosity measured in the proper frame of the jet as $L_{\gamma, em} \simeq L_{\gamma}/2\Gamma^{2} \simeq$~1.2 $\times$ 10$^{46}$ erg s$^{-1}$ considering a bulk Lorentz factor $\Gamma$ = 20 obtained from the SED modeling (Table~\ref{tab:sed}). This is a good fraction of the kinetic jet power ($\sim$~53\%; ${\rm P}_{j, kin}$ = 2.2 $\times$ 10$^{46}$ erg s$^{-1}$) indicating that the jet becomes radiatively efficient and a significant amount of the kinetic jet power gets converted to the radiative power.

The SEDs of J0809+5341 reveal that the optical-UV spectrum is steep which we interpret as synchrotron emission. Alternatively, one can associate this spectrum to the accretion disk emission. However, it is unlikely that the large variation seen at optical-UV energies during low and high activity states can be a result of the perturbations in the accretion disk emission. Moreover, the analysis of the SDSS spectrum also suggests a relatively less luminous disk (Section~\ref{subsec:bhmass}). In addition, the source has shown high optical polarization during the recent flare, thus supporting the synchrotron origin of the optical-UV spectrum. The X-ray spectrum of the source, obtained from the {\it Swift}-XRT and {\it NuSTAR} observations during the flare, is typical of powerful blazars and can be well reproduced by the SSC process.

\begin{figure*}

\hbox{\hspace{-1.5cm}
      \includegraphics[width=10.0cm]{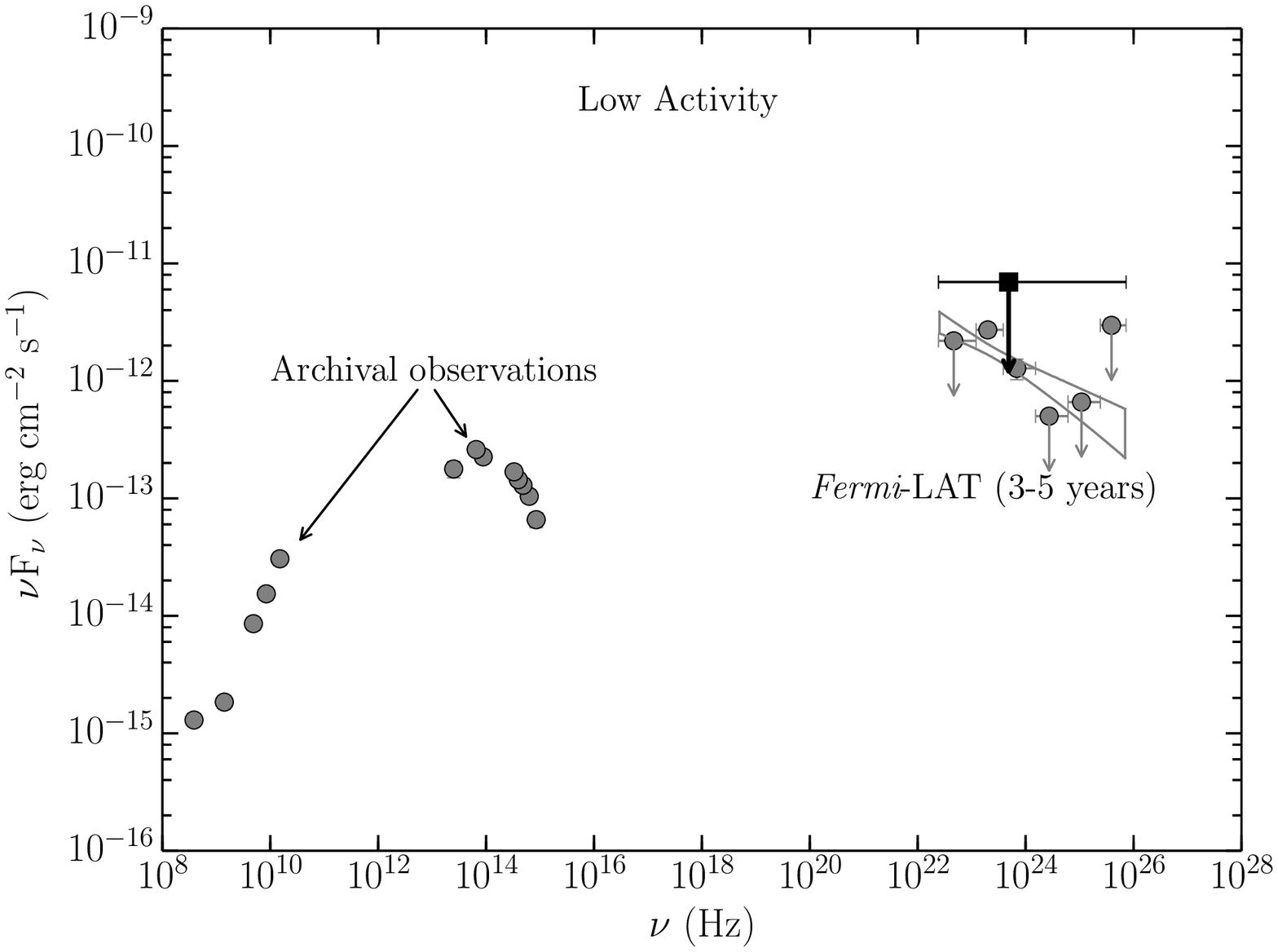}
      \includegraphics[width=10.0cm]{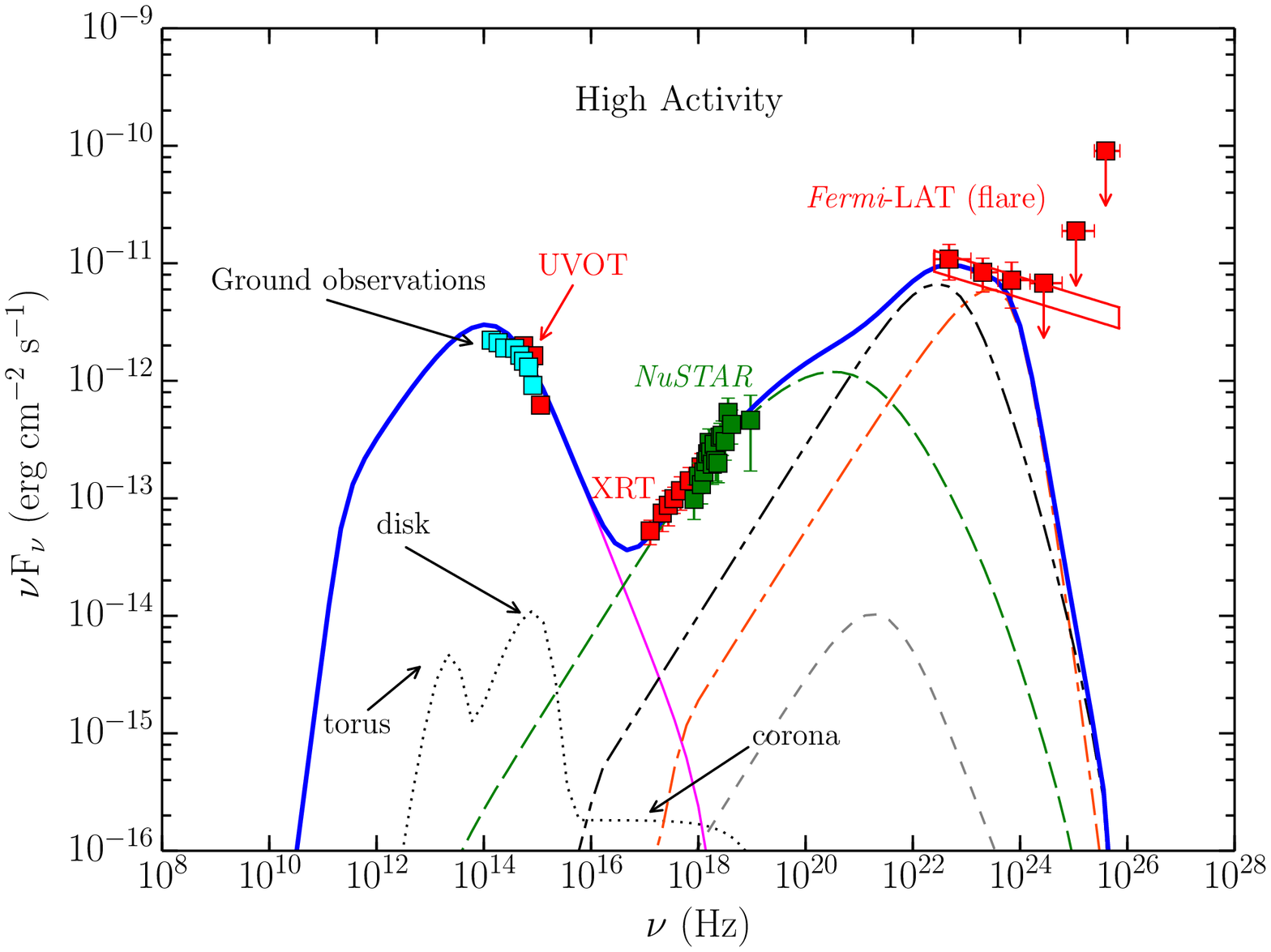}
     }
\hbox{\hspace{3cm}
      \includegraphics[width=10.0cm]{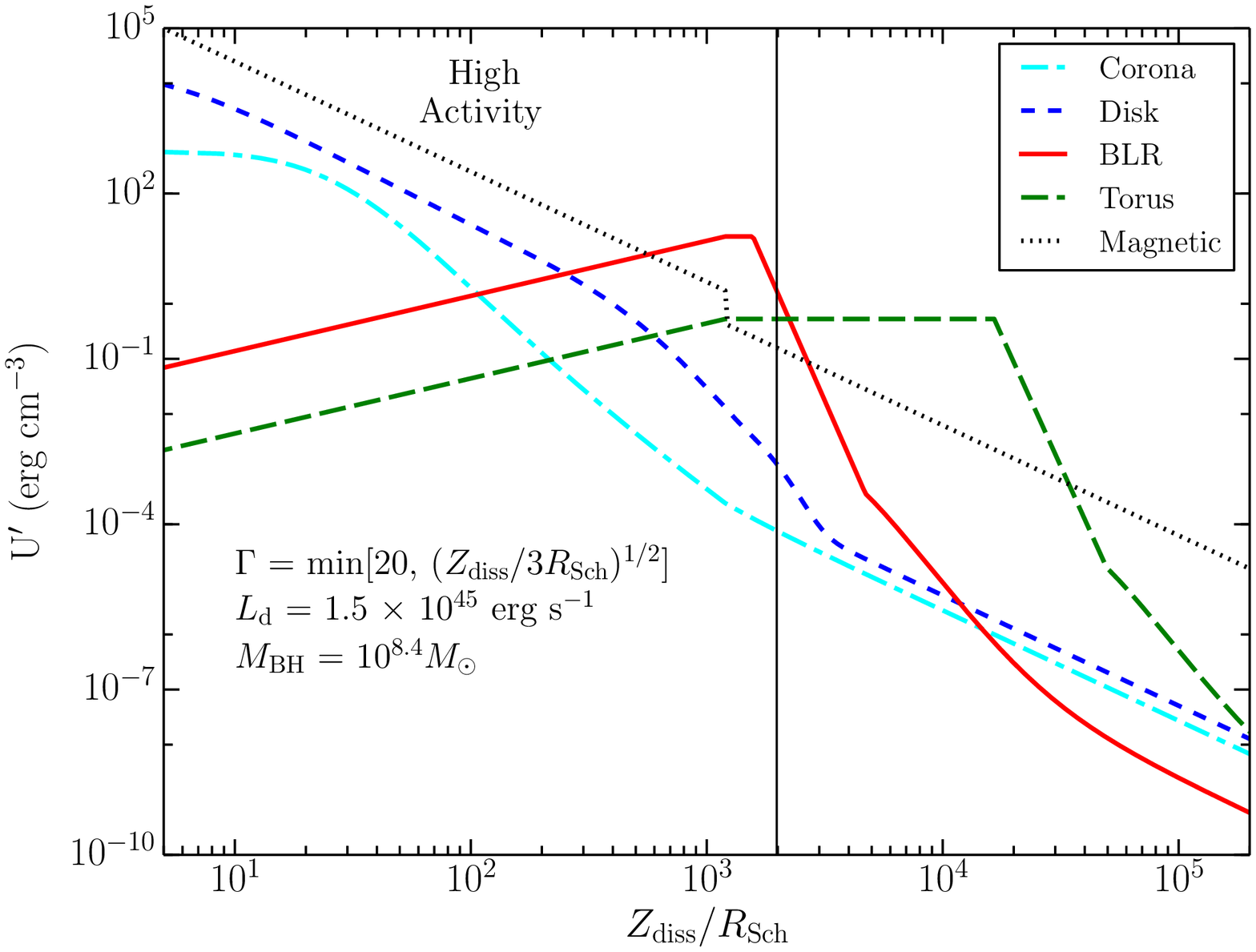}
     }
\caption{Top left: Low activity state SED of J0809+5341. Gray bow-tie plot represents third through fifth year of \fermi~operation. Black downward arrow is the first two years 95\% flux upper limit, calculated by assuming the photon index obtained from the analysis of the third through fifth year LAT data. Top right: Modeled high activity state SED. Black dotted line represents thermal contributions from the torus, accretion disk, and X-ray corona. Pink thin solid line and green long dashed line are synchrotron and SSC radiation. Grey dashed, red dash-dot, and black dash-dot-dot lines represent EC-disk, EC-BLR, and EC-torus components respectively. Blue thick solid line refers to the sum of all the radiation components. Bottom panel: Variation of the energy densities measured in the comoving frame, as a function of the distance from the black hole, in units of $R_{\rm Sch}$. Vertical line denotes the location of the emission region.}\label{fig:sed}
\end{figure*}

It can be seen in Figure~\ref{fig:sed} that the shape of the $\gamma$-ray spectrum is relatively flat compared to the optical-UV spectrum. A steep optical-UV spectrum suggests that the spectral shape of the electron energy distribution is soft. If electrons of similar energies are contributing to the emission at $\gamma$-rays through IC process, then one would expect a steep $\gamma$-ray spectrum similarly to the optical-UV. However, we observed a hard $\gamma$-ray spectrum. Hence, we assume that the harder $\gamma$-ray spectrum is the result of the superposition of different IC emission processes. One possibility is that the SSC is contributing in the hard X-ray to soft $\gamma$-rays and an EC component explaining the remaining spectra, however, this is not possible since the steep falling optical spectrum suggests that the synchrotron emission peaks at lower energies which in turn causes negligible SSC emission at high energies \citep[e.g.,][]{2012MNRAS.419.1660S}. This indicates that, though significant, SSC cannot explain the observed $\gamma$-ray spectrum. Alternatively, a flat $\gamma$-ray spectrum can be a result of interplay between various EC mechanisms. We find that the inclusion of EC-BLR and EC-torus emission can reproduce the $\gamma$-ray spectrum satisfactorily. The relative contribution of these target photons measured in the comoving frame depends on the location of the emission region from the central engine. Then, under the assumption that the $\gamma$-ray spectrum is the superposition of EC-BLR and EC-torus components, we can derive the location of the emission region by reproducing the observed $\gamma$-ray spectral shape. In the bottom panel of Figure~\ref{fig:sed}, we plot the energy densities of various components in the comoving frame, as a function of $Z_{\rm diss}$. Modeling the flaring SED indicates that the emission region is located at a distance from the central black hole where both the BLR and IR-torus energy densities are contributing almost equally to produce the observed $\gamma$-ray spectrum. We further constrain the parameters of the present model by considering near-equipartition between relativistic particles and magnetic field. The resulting model spectrum along with the observed fluxes are shown in Figure~\ref{fig:sed} and the corresponding parameters are given in Table~\ref{tab:sed}.

Comparison of the flaring state SED with that representing the low activity state indicates few interesting features. The increase of the optical flux appears to be higher than that of the $\gamma$-ray flux. This is supported by the fact that, at the time of the flare, the source was one of the brightest quasars in the optical band whereas the rise of the $\gamma$-ray flux was relatively modest. However, it should be noted that we do not have simultaneous $\gamma$-ray observations at the time of low activity state as recorded by the SDSS optical monitoring and thus a strong claim by comparing the non-simultaneous observations cannot be made. Further, comparison of the accretion disk flux with that of the archival optical observations hints that even during quiescence the optical-UV emission is synchrotron dominated.

The SEDs and associated modeling parameters of J0809+5341 are quite different from those obtained for other high redshift blazars. Generally, the optical-UV part of the SED of high redshift blazars is dominated by extremely luminous accretion disk radiation, the peak of IC emission lies in the hard X-ray regime resulting in a steep $\gamma$-ray spectrum, and they also are known to host more than a billion solar mass black hole at their centers \citep[e.g.,][]{2011MNRAS.411..901G,2013MNRAS.428.1449G}. In contrast, J0809+5341 hosts a relatively less luminous accretion disk and less massive central black hole. The optical-UV spectrum is dominated by synchrotron radiation, and the IC peak lies at GeV range.  Thus, the overall observed properties of J0809+5341 indicates that this source is, in many ways, different from other high redshift blazars but show similarities with its low redshift counterparts \citep{2010MNRAS.402..497G}.

\section{Conclusions}\label{sec:conclu}
In this work, we present a detailed multi-frequency study of the high redshift blazar J0809+5341. The main findings of the work are summarized below.
\begin{enumerate}
 \item Predicted as a candidate $\gamma$-ray emitter by \citet{2008ApJS..175...97H}, J0809+5341 is now detected in the $\gamma$-ray band by \fermi-LAT, as confirmed by the 3FGL catalog.
 \item The black hole mass and accretion disk luminosity, calculated by analyzing the archival SDSS spectrum, are found to be 10$^{8.4} M_{\odot}$ and 1.5 $\times$ 10$^{45}$ erg s$^{-1}$ respectively.
 \item A significant fraction of the kinetic jet power gets converted to the radiative power during the flare, or in other words, the jet becomes radiatively efficient.
 \item The flaring state optical-UV spectrum can be successfully modeled by the synchrotron emission. The choice of the synchrotron mechanism over the accretion disk is primarily influenced by the observation of high optical polarization during the flare and by the flare itself.
  \item The optical-UV spectrum is found to be steep and the $\gamma$-ray spectrum is relatively flat. The flatness of the $\gamma$-ray spectrum can be explained by locating the emission region outside the BLR where both the BLR and torus energy densities play a major role in describing the observed $\gamma$-ray spectrum. 
 \item Many of the observed properties of J0809+5341 are at odds with that generally observed in other high redshift blazars \citep{2011MNRAS.411..901G,2013MNRAS.428.1449G}.
\end{enumerate}

\acknowledgments
We thank the referee for a constructive report that improved the manuscript significantly. We are thankful to the {\it Swift} PI for accepting the ToO request and the {\it Swift} operation duty scientist for quickly scheduling the observations. We also thank {\it NuSTAR} PI for approving the request of observation. VSP is grateful to MASTER-Tunka team for providing the optical images of the source. This research has made use of the data obtained from HEASARC provided by NASA’s Goddard Space Flight Center. Part of this work is based on archival data, software, or online services provided by the ASI Science Data Center (ASDC). This research has made use of the XRT Data Analysis Software (XRTDAS) developed under the responsibility of the ASDC, Italy. This research has also made use of the NuSTAR Data Analysis Software (NuSTARDAS) jointly developed by the ASI Science Data Center (ASDC, Italy) and the California Institute of Technology (Caltech, USA).

Funding for SDSS-III has been provided by the Alfred P. Sloan Foundation, the Participating Institutions, the National Science Foundation, and the U.S. Department of Energy Office of Science. The SDSS-III web site is http://www.sdss3.org/.

SDSS-III is managed by the Astrophysical Research Consortium for the Participating Institutions of the SDSS-III Collaboration including the University of Arizona, the Brazilian Participation Group, Brookhaven National Laboratory, Carnegie Mellon University, University of Florida, the
French Participation Group, the German Participation Group, Harvard University, the Instituto de Astrofisica de Canarias, the Michigan State/Notre Dame/JINA Participation Group, Johns Hopkins University, Lawrence Berkeley National Laboratory, Max Planck Institute for Astrophysics, Max Planck Institute for Extraterrestrial Physics, New Mexico State University, New York University, Ohio State University, Pennsylvania State University, University of Portsmouth, Princeton University, the Spanish Participation Group, University of Tokyo, University of Utah, Vanderbilt University, University of Virginia, University of Washington, and Yale University.

Use of {\it Hydra} cluster at the Indian Institute of Astrophysics is acknowledged.

\bibliographystyle{apj}
\bibliography{Master}

\begin{table}
\caption{Details of the power law model fits to the averaged $\gamma$-ray data for various time periods. The quoted flux values are in units of 10$^{-8}$ \phflux~whereas $L_{\gamma}$ is the $\gamma$-ray luminosity. The last column quotes the significance of the curvature present in the spectrum by means of a LogParabola model fitting.}
\begin{center}
%\begin{flushleft}
 \begin{tabular}{cccccc}
\hline \hline
Time Period (MJD) & $\Gamma_{0.1-300~{\rm GeV}}$ & $F_{0.1-300~{\rm GeV}}$ &  TS &log ${L_{\gamma}}$ & ${TS_{\rm curve}}$ \\
\hline
54683$-$55412 (2008 Aug 05 to 2010 Aug 4) & -- & --  &  0.5 & -- & -- \\
55412$-$56508 (2010 Aug 04 to 2013 Aug 4) & 2.26 $\pm$ 0.08 & 1.61 $\pm$ 0.26  & 185.5 & 47.72 & 8.70\\
56508$-$56873 (2013 Aug 04 to 2014 Aug 4) & 2.15 $\pm$ 0.08 & 2.78 $\pm$ 0.44  & 191.3 & 48.00 & 9.64\\
54683$-$56873 (2008 Aug 05 to 2014 Aug 4) & 2.28 $\pm$ 0.06 & 1.40 $\pm$ 0.19  & 267.6 & 47.65 & 9.60\\
\hline
\end{tabular}
\label{tab:fermi_lat}
\\
\end{center}
%\end{flushleft}
\end{table}

\begin{table}
\caption{{\it Swift}-UVOT observations of J0809+5341. The quoted numbers are in magnitudes.}
\begin{center}
\begin{tabular}{ccccc}
\hline\hline
UVOT-filters & 2014 April 26 & 2014 April 27 &  2014 April 28 & 2014 May 8 \\
\hline
UVW1  & --               & 19.17 $\pm$ 0.12  &  --               & 18.62 $\pm$ 0.07 \\
U     & 17.39 $\pm$ 0.08 & --                & 17.93 $\pm$ 0.09  & 17.28 $\pm$ 0.04 \\
B     & 18.30 $\pm$ 0.14 & --                & 18.69 $\pm$ 0.11  & 18.00 $\pm$ 0.05 \\
V     & 17.58 $\pm$ 0.05 & --                & 17.97 $\pm$ 0.13  & 17.42 $\pm$ 0.07 \\
\hline
\end{tabular}
\label{tab:uvot_phot}
\end{center}
\end{table}

\begin{table}
\caption{Apparent magnitudes of J0809+5341, as observed from HCT. For comparison, the ugriz magnitudes from SDSS that are converted to UBVRI magnitudes are also given.}
\begin{center}
\begin{tabular}{cccc}
\hline\hline
Filters & 2014 April 27 & 2014 May 1 &  2003 November 20 (SDSS) \\
\hline
U & --               & $18.15 \pm 0.07$  &  $21.03 \pm 0.12$  \\
B & $18.14 \pm 0.04$ & $18.48 \pm 0.04$  &  $21.19 \pm 0.04$ \\
V & $17.67 \pm 0.02$ & $17.99 \pm 0.03$  &  $20.61 \pm 0.04$ \\
R & $17.17 \pm 0.01$ & $17.47 \pm 0.03$  &  $20.21 \pm 0.04$ \\
I & $16.56 \pm 0.02$ & $16.82 \pm 0.04$  &  $19.69 \pm 0.05$  \\
\hline
\end{tabular}
\label{tab:hct_phot}
\end{center}
\end{table}

\begin{table*}
\begin{center}
{
\small
\caption{Summary of the SED analysis\label{tab:sed_par}}
\begin{tabular}{cccccc}
\tableline\tableline
 & & {\it Fermi}-LAT & &  & \\
\tableline
 Activity state\tablenotemark{a} & Period\tablenotemark{b} & $F_{0.1-300~{\rm GeV}}$\tablenotemark{c} & $\Gamma_{0.1-300~{\rm GeV}}$\tablenotemark{d} & TS\tablenotemark{e} & {\it N$_{\rm pred}$}\tablenotemark{f}  \\
 \tableline
 Q & 55412$-$56508  &1.61~$\pm$~0.26 & 2.26~$\pm$~0.08 & 185.47 & 897.63  \\
 F & 56748$-$56786  &5.87~$\pm$~1.19 & 2.14~$\pm$~0.08 & 50.48  & 110.94 \\
 \tableline
 & & {\it NuSTAR}  & & &\\
 \tableline
 Activity state & Exp.\tablenotemark{g}& $\Gamma_{3-79~{\rm keV}}$\tablenotemark{h} & $F_{3-79~{\rm keV}}$\tablenotemark{i} & Normalization\tablenotemark{j}& Stat.\tablenotemark{k}\\
\tableline
 F & 28 & 1.30~$\pm$~0.30  & 1.58$^{+0.65}_{-0.46}$ & 3.30~$\pm$~0.40 & 12/14\\
 \tableline
 & & {\it Swift}-XRT  & & &\\
 \tableline
 Activity state & Exp.\tablenotemark{g}& $\Gamma_{0.3-10~{\rm keV}}$\tablenotemark{l} & $F_{0.3-10~{\rm keV}}$\tablenotemark{m} & Normalization\tablenotemark{j}& Stat.\tablenotemark{k}\\
\tableline
 F & 9.45 & 1.45$^{+0.30}_{-0.29}$ & 4.51$^{+1.32}_{-1.00}$ & 5.09$^{+1.29}_{-1.13}$ & 52.14/74\\
 \tableline
 & & {\it Swift}-UVOT &  & & \\
 \tableline
 Activity state & V\tablenotemark{n}& B\tablenotemark{n}& U\tablenotemark{n}& UVW1\tablenotemark{n}& \\
 \tableline
 F & 1.99~$\pm$~0.09 & 1.56~$\pm$~0.09 & 1.64~$\pm$~0.06 & 0.62~$\pm$~0.02 & \\
 \tableline
\end{tabular}
\tablenotetext{1}{Different activity states selected for the modeling; Q: low activity and F: high activity state.}
\tablenotetext{2}{Time period considered for the SED modeling, in MJD.}
\tablenotetext{3}{Integrated $\gamma$-ray flux in 0.1$-$300 GeV energy range in units of 10$^{-8}$ \phflux.}
\tablenotetext{4}{Photon index calculated from $\gamma$-ray analysis.}
\tablenotetext{5}{Significance of detection using likelihood analysis.}
\tablenotetext{6}{Number of predicted photons during the time period under consideration.}
\tablenotetext{7}{Net exposure in kiloseconds.}
\tablenotetext{8}{Photon index of the power law model.}
\tablenotetext{9}{Observed flux in units of 10$^{-12}$ erg cm$^{-2}$ s$^{-1}$, in 3$-$79 keV energy band.}
\tablenotetext{10}{Normalization at 1 keV in 10$^{-5}$ ph cm$^{-2}$ s$^{-1}$ keV$^{-1}$.}
\tablenotetext{11}{Statistical parameters: C-stat./dof.}
\tablenotetext{12}{Photon index of the absorbed power law model.}
\tablenotetext{13}{Unabsorbed flux in units of 10$^{-13}$ erg cm$^{-2}$ s$^{-1}$, in 0.3$-$10 keV energy band.}
\tablenotetext{14}{Average flux in {\it Swift} V, B, U, and UVW1 bands, in units of 10$^{-12}$ erg cm$^{-2}$ s$^{-1}$.}
%\tablecomments{See the text for details.}
}
\end{center}
\end{table*}

\begin{table*}
{\small
\begin{center}
\caption{Summary of the parameters used/derived from the modeling of the SED.}\label{tab:sed}
\begin{tabular}{lcc}
\tableline
\tableline
Parameter                                            &  Symbol                &   High Activity  \\
\tableline
Slope of particle spectral index before break energy & {\it p}                & 1.6    \\
Slope of particle spectral index after break energy  & {\it q}                & 5.3    \\
Magnetic field in Gauss                              & {$B$}                  & 2.0    \\
Particle energy density in erg cm$^{-3}$             & $U'_{\rm e}$           & 0.05   \\
Bulk Lorentz factor                                  &$\Gamma$                & 20     \\
Minimum Lorentz factor                               & $\gamma'_{min}$        & 1      \\
Break Lorentz factor                                 & $\gamma'_{b}$          & 1714   \\
Maximum Lorentz factor                               & $\gamma'_{max}$        & 1e5    \\
Size of the BLR in the units of $R_{\rm Sch}$        & $Z_{\rm BLR}$          & 1570    \\
Distance of the emission region from the black hole in parsec ($R_{\rm Sch}$) & $Z_{\rm diss}$ & 0.048(1980)\\
Dusty torus temperature in Kelvin                       & $T_{\rm IR}$        & 900    \\
Viewing angle in degrees                             & $\theta$               & 3       \\
\hline
Jet power in electrons in log scale                  & $P_{\rm e}$            & 44.62  \\
Jet power in magnetic field in log scale             & $P_{\rm B}$            & 45.11  \\
Radiative jet power in log scale                     & $P_{\rm r}$            & 46.19  \\
Jet power in protons in log scale                    & $P_{\rm p}$            & 46.34  \\
\tableline
\end{tabular}
\end{center}
}
\end{table*}

\newpage

\newpage

\newpage

\end{document}